%% file: paper.tex
\documentclass[conference]{IEEEtran}

\newif\ifanon
\anonfalse

\usepackage{amsmath,amsfonts}
\usepackage{algorithmic}
\usepackage{graphicx}
\usepackage{textcomp}
\usepackage{xcolor}
\usepackage{booktabs}
\usepackage{todonotes}
\usepackage{tabularx}
\usepackage{makecell}
\usepackage{listings}
\usepackage{xcolor}

\usepackage[export]{adjustbox}

\usepackage{subcaption}

\usepackage{cleveref}
\usepackage{xspace}

\usepackage[T1]{fontenc}
\usepackage[utf8]{inputenc}

\definecolor{codegreen}{rgb}{0,0.6,0}
\definecolor{codegray}{rgb}{0.5,0.5,0.5}
\definecolor{codepurple}{rgb}{0.58,0,0.82}
\definecolor{backcolour}{rgb}{0.95,0.95,0.92}

\lstdefinestyle{mystyle}{
    backgroundcolor=\color{backcolour},
    commentstyle=\color{codegreen},
    keywordstyle=\color{magenta},
    numberstyle=\tiny\color{codegray},
    stringstyle=\color{codepurple},
    basicstyle=\ttfamily\footnotesize,
    breakatwhitespace=false,
    breaklines=true,
    captionpos=b,
    keepspaces=true,
    numbers=left,
    numbersep=5pt,
    showspaces=false,
    showstringspaces=false,
    showtabs=false,
    tabsize=2
}

\newcolumntype{L}{>{\raggedright\arraybackslash}X}
\newcolumntype{C}{>{\centering\arraybackslash}X}
\newcolumntype{R}{>{\raggedleft\arraybackslash}X}
\newcolumntype{s}{>{\hsize=.5\hsize}C}
\def\BibTeX{{\rm B\kern-.05em{\sc i\kern-.025em b}\kern-.08em
    T\kern-.1667em\lower.7ex\hbox{E}\kern-.125emX}}

\ifanon
    \newcommand{\toolname}{Akira\xspace}
    \newcommand{\toolnamelong}{Agentic Kernel Inspection, Reasoning, and Analysis tool\xspace}
\else
    \newcommand{\toolname}{KEET\xspace}
    \newcommand{\toolnamelong}{Kernel Execution Explanation Tool\xspace}
\fi

\begin{document}

\ifanon
    \title{Automated Interpretation of GPU Profiling Data Using Agentic LLM Systems}
\else
    \title{\toolname: Explaining Performance of GPU Kernels Using LLM Agents}
\fi

\ifanon
    \author{\IEEEauthorblockN{Anonymous Author(s)}}
\else
    \author{
        \IEEEauthorblockN{Joshua H. Davis\IEEEauthorrefmark{2}, Klaudiusz Rydzy\IEEEauthorrefmark{2}, Srinivasan Ramesh\IEEEauthorrefmark{1}, Aadit Nilay\IEEEauthorrefmark{2}, Daniel Nichols\IEEEauthorrefmark{3},\\ Swapna Raj\IEEEauthorrefmark{1}, Nikhil Jain\IEEEauthorrefmark{1}, Abhinav Bhatele\IEEEauthorrefmark{2}}
        ~\\
        \IEEEauthorblockA{
           \IEEEauthorrefmark{2}Department of Computer Science, University of Maryland, College Park, MD, USA\\
           \IEEEauthorrefmark{1}NVIDIA, Inc., Santa Clara, CA, USA\\
           \IEEEauthorrefmark{3}Center for Applied Scientific Computing, Lawrence Livermore National Laboratory, Livermore, CA, USA\\
           jhdavis@umd.edu, bhatele@cs.umd.edu
        }
    }
\fi

\maketitle

\begin{abstract}
    \input{abstract}
\end{abstract}

\section{Introduction}
\label{sec:introduction}
\input{intro}

\section{Background}
\label{sec:background}
\input{bg}

\section{Design of \toolname}
\label{sec:keet}
\input{keet}

\section{Methodology for Evaluating Explanations}
\label{sec:method}
\input{method}

\section{Evaluation Setup}
\label{sec:setup}
\input{setup}

\section{Results and Discussion}
\label{sec:results}
\input{results}

\section{Related Work}
\label{sec:related}
\input{related}

\section{Conclusion}
\label{sec:conclusion}
\input{conc}

\bibliographystyle{IEEEtran}
\bibliography{bib/pssg,bib/cite}

\end{document}

%% file: abstract.tex
Performance profiles of GPU kernels generated by tools such as Nsight Compute
are rich in detail but are often challenging to interpret. To achieve the best
performance possible on a given GPU architecture, kernel developers need to
spend significant time analyzing and comparing profiles in the tool's graphical
interface to identify and understand kernel performance bottlenecks. Large
Language Models (LLMs) have shown promise in understanding complex data and
generating natural language explanations.  In this paper, we propose the
\toolnamelong (\toolname), an LLM-based agentic framework for interpreting
Nsight Compute profiles to generate useful and data-grounded natural language
explanations of performance issues in GPU kernels, and suggestions for
optimizations. We evaluate \toolname using several CUDA kernels of varying
complexity on NVIDIA H100 GPUs. We find that the generated explanations, when
provided as context, improve the quality of LLM code optimization and
multiple-choice question answering in downstream tasks. We further demonstrate
that the tool can be used to interpret performance data from large sets of
profiles to improve the quality of optimization suggestions.

%% file: intro.tex
Graphics Processing Units (GPUs) have become ubiquitous in modern
supercomputers and clusters. The Top500 list~\cite{top500nov2025}, as of
November 2025, includes 255 systems (51\% of the full list) that use GPUs or
other accelerators. Further, nine out of the top ten systems use GPUs.
Scientific applications must utilize GPUs as efficiently as possible to achieve
maximum performance on these systems. Writing and optimizing code to achieve
this goal on a given GPU architecture requires careful consideration of the
hardware architecture and the requirements of the particular application.
Meanwhile, modern GPU architectures are becoming increasingly sophisticated.
Each new generation adds not just additional streaming multiprocessors and
memory bandwidth, but also new features, such as specialized cores and more
complex memory hierarchies. In the worst-case scenario, every GPU architecture
release requires a new profiling and optimization effort to achieve peak
performance, which is an unacceptable drain on developer time and resources.

GPU software developers and performance engineers employ profiling tools such
as NVIDIA's Nsight Compute (NCU)~\cite{ncu} to understand and optimize the
performance of their code on a given GPU architecture. In particular,
developers must identify which hardware components and code regions are causing
performance bottlenecks. However, despite the wealth of information these tools
provide, extracting insight from GPU profiling tools remains a complex and
largely manual task. While NCU provides a graphical interface to visualize
metrics and a rule engine to automatically highlight potential problems, it
remains challenging to easily and quickly identify key insights, especially
from large sets of profiles. In this work, we address the challenge of
extracting actionable and informative optimization insights from individual and
multiple GPU kernel profiles.

Tools such as GPA~\cite{zhou2021gpa} and DrGPU~\cite{hao2023drgpu} have
proposed automated approaches to assist with analyzing GPU kernel performance
on NVIDIA GPUs, but both these tools take a rule-based approach that can be
somewhat rigid. DrGPU, for example, decomposes stall reasons into categories in
a tree structure and offers suggestions to address the most frequent stall
reasons. However, these suggestions are statically defined and do not consider
how specific characteristics of the kernel algorithm might interact with
performance bottlenecks. Further, these existing tools require manual
intervention to incorporate new GPU hardware features or performance metrics
when new GPUs are released. Most importantly, they do not suggest code changes
tailored to the existing kernel code.

We propose a new approach to address these limitations by interpreting GPU
kernel performance profiles using Large Language Models (LLMs). Our
\toolnamelong (\toolname) is an LLM-based agentic framework that automatically
generates a natural language report explaining the performance of a GPU kernel
using NCU profiles. It identifies key performance bottlenecks and suggests
specific code changes to address them. Further, it scales easily to larger sets
of profiles and can be used to understand the relationship of kernel tuning
knobs such as block size or algorithm parameters to performance and hardware
behavior.

By using LLMs to generate these reports, \toolname provides natural language
explanations grounded in the measured profiling data, rather than reciting data
or results of static rules. We leverage public knowledge of GPU architectures,
NCU performance metrics, and optimization strategies embedded in pre-trained
LLMs to generate useful insights from profiling metrics and kernel code.
Because the approach does not rely on statically defined rules, and instead
uses pre-trained LLMs as the core reasoning system, it has the potential to be
easily updated to understand newer GPU architectures by updating to a model
with a more recent knowledge frontier. In this work, we present and describe
the design of \toolname, evaluate its effectiveness against prior work, and
study its sensitivity to input profile dataset size as well as the contribution
of individual LLM agent roles to overall output quality.

Our work makes the following novel contributions:
\begin{itemize}
    \item We present \toolname, an LLM-based agentic tool that automatically
generates a natural language report explaining the performance of a CUDA kernel
profiled with Nsight Compute, identifying key performance bottlenecks, and
suggesting specific code changes to address them.
    \item We evaluate the quality of \toolname's analysis of GPU kernels
against DrGPU using two downstream tasks -- multiple-choice question answering
and code optimization -- over several CUDA kernels of varying complexity. We
find that \toolname's explanations enable LLMs to answer more questions
correctly and implement stronger optimizations compared to DrGPU's output.
    \item We demonstrate that \toolname can be used to interpret performance
data from large sets of profiles to understand the relationship of kernel
tuning knobs to performance, and study how the quality of explanations improves
with additional profiles provided to the tool.
    \item We ablate the agent roles included in \toolname to understand their
individual impact on output quality.
\end{itemize}

%% file: bg.tex
Below we provide some background details on the Nsight Compute (NCU)
profiling tool, metrics for evaluating LLM-generated GPU code, and the
DrGPU tool that we optionally incorporate into our framework.

\subsection{Nsight Compute Profiler}
\label{sec:background:ncu}

Nsight Compute (NCU) is a profiling tool for NVIDIA GPUs that collects detailed
performance metrics at the level of individual GPU kernels, down to the source
code line level if requested. NCU allows users to specify which kernels in the
application to profile and which metrics to collect. It profiles targeted
kernels by replaying each kernel for multiple passes, collecting different
hardware metrics on each pass. NCU determines the number of passes required
based on the set of metrics specified by the user.

After profiling, NCU generates a profile report file (\texttt{.ncu-rep})
containing the collected metrics for each kernel. The report file can be
viewed in the NCU graphical interface, which provides a variety of views of
the collected data. These views include a summary view of the basic details
of each kernel profiled, a detailed view of the metrics collected for a
selected kernel, grouped into sections roughly by hardware component, and a
source code view that maps metrics to source code and assembly lines.
Notably, the detailed view also provides outputs from a rule engine that
generates callouts mentioning potential performance issues and suggestions
based on fixed metric thresholds.

Data generated by NCU is typically viewed through the graphical interface,
but it can also be extracted from the \texttt{.ncu-rep} file using NCU's
Python Report Interface (PRI). PRI allows for programmatically viewing
collected metric names, values, units, and other metadata for each kernel
profiled in the report file. Importantly, PRI does not provide full access
to line-level profiling data, which must be manually exported to a CSV file
through the graphical interface. We primarily use PRI to extract the raw
metric data for each kernel profiled for use in our tool's analysis. DrGPU,
as described in Section~\ref{sec:background:drgpu}, also requires the
line-level data to produce the best results. We manually export the
required CSV file to support DrGPU's analysis in both baseline and
\toolname evaluations.

\subsection{Evaluating LLM-generated Code}
\label{sec:background:metrics}

As described in Section~\ref{sec:method:downstream-tasks}, we evaluate the
quality of \toolname's analysis of GPU kernel performance data using two
baseline tasks. One of these, the optimization (OPT) task, asks the LLM to
generate an optimized version of the kernel code based on the provided
analysis. LLM attempts to generate performance-optimized code in HPC contexts
have been widely studied~\cite{nichols:hpdc2024, godoy2024large,
valero2024chatblas}. We rely on two existing metrics proposed in prior work:
pass@k and speedup@k~\cite{nichols:hpdc2024}.

The pass@k score estimates the fraction of code generation attempts that
generate valid (compilable and test case-passing) code given k attempts. For
$N$ samples, where $N>k$, the pass@k score is given by \Cref{eq:pass-k}. We
denote the number of correct samples by $c_t$ for a task $t$ of $T$ total
tasks.

\vspace{1.5em}
\begin{equation}\label{eq:pass-k}
    \text{pass@}k =
    \frac{1}{\lvert T\rvert}
    \sum_{t\in T}
    \left[
        1 -
        \binom{
            N -
            c_t
        }{
            k
        }
        /
        \binom{
            N
        }{
            k
        }
    \right]
\end{equation}
\vspace{1.5em}

\begin{figure*}[ht]
    \centering
    \ifanon
        \includegraphics[width=0.9\textwidth]{figs/Akira.pdf}
    \else
        \includegraphics[width=0.9\textwidth]{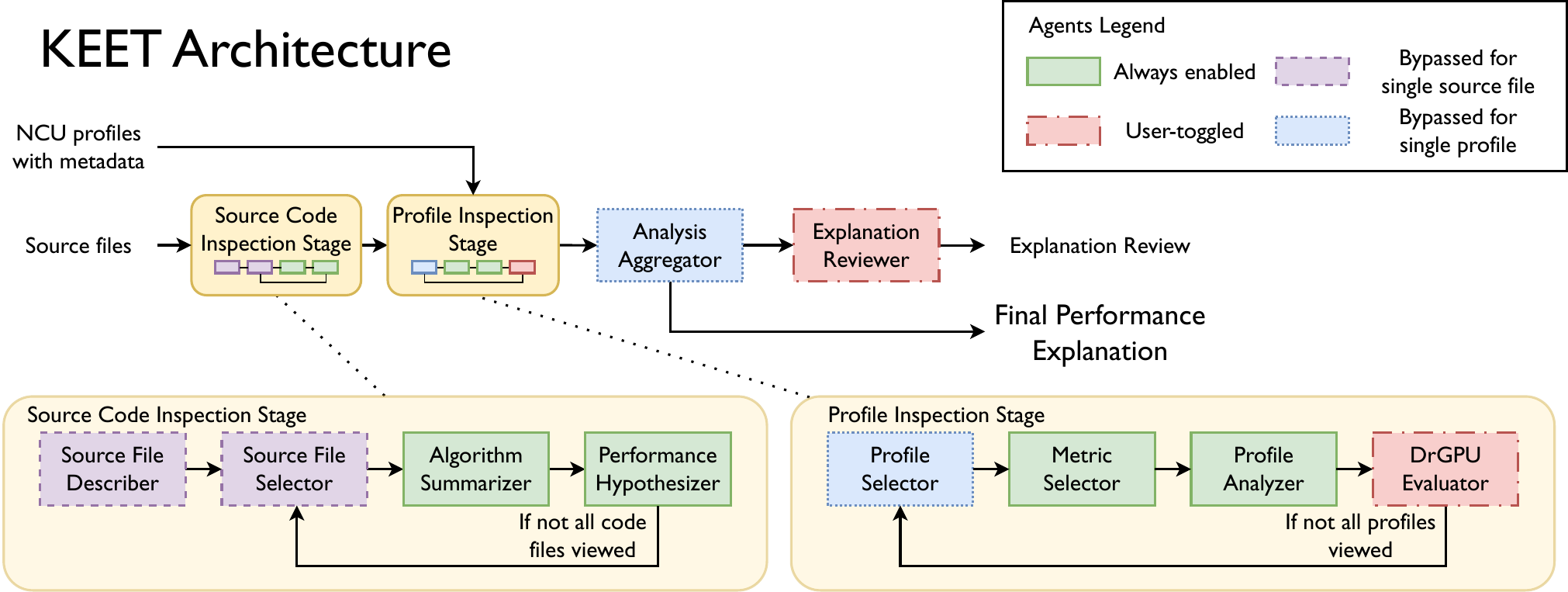}
    \fi
    \caption{The architecture of \toolname, where each green box represents
    an agent role and the arrows represent the data flow between them. The
    Source Code Inspection and Profile Inspection stages are depicted in
    detail below the overall architecture diagram.}
    \label{fig:tool-architecture}
\end{figure*}

The speedup@k score is conceptually similar to pass@k, but estimates the
expected speedup in performance over the original (unoptimized) code achieved
with $k$ attempts. For $N$ samples, where $N>k$, the speedup@k score is
calculated as \Cref{eq:speedup-k}. We denote the performance of the original
code by $T^*_p$, and the performance of the updated code on sample $j$ by
$T_{p,j}$, where $p$ is a problem from a set of $P$ problems.

\begin{equation}\label{eq:speedup-k}
    \textrm{speedup}@k =
    \frac{1}{\lvert P\rvert}
    \sum_{p\in P}
    \sum_{j=1}^N
    \frac{
        \binom{
            j-1
        }{
            k-1
        }
    }{
        \binom{
            N
        }{
            k
        }
    }
    \frac{
        T^*_p
    }{
        T_{p,j}
    }
\end{equation}

We evaluate the quality of LLM-generated code using pass@k and speedup@k with
$N=20$ samples and $k=1$.

\subsection{DrGPU Profile Analysis Tool}
\label{sec:background:drgpu}

DrGPU is a non-LLM-based tool that generates performance reports and
optimization suggestions based on an NCU profile. It examines kernel
performance by decomposing stall reasons into categories in a tree
structure, listing root causes at leaf nodes. DrGPU offers suggestions to
address the most frequent stall reasons on these leaf nodes. The tool
requires a specific set of NCU metrics to run its analysis. Furthermore, in
order to provide source code lines alongside its suggestions, it also
requires the user to manually export a CSV file containing the line-level
profiling data, as described in Section~\ref{sec:background:ncu}. In this
paper we compare \toolname against DrGPU as a baseline and optionally
integrate its suggestions into our tool's analysis.

%% file: keet.tex
\toolname is an LLM-based agentic framework for explaining the performance of
GPU kernels by analyzing their Nsight Compute (NCU) performance profiles. Its
design emphasizes scalability to large numbers of profiles, interpretability of
intermediate analysis steps, and integration of algorithmic understanding.
\toolname takes as input a single NCU profile, or a set of multiple profiles,
along with the kernel source code and descriptions of the run configuration(s)
used to generate the profile(s). It produces a natural language report that
explains kernel performance, identifies key code locations and hardware
bottlenecks, analyzes the relationship between tuning knobs and performance,
and suggests specific optimizations.

\subsection{Overview of \toolname}

We present the overall structure of \toolname in
Figure~\ref{fig:tool-architecture}. Each colored box represents an LLM agent
role -- a prompt template populated with appropriate context from prior agents
as indicated by arrows. As indicated in the legend, some agent roles are
optional, and others are bypassed depending on whether handling for multiple
source files or multiple profiles is required. The tool takes as input the
kernel source code files, NCU profile(s), and descriptive metadata. It outputs
the final performance explanation report and an accompanying review.

Rather than relying on a single monolithic LLM prompt, \toolname decomposes
analysis into multiple specialized agent roles. Together, these agents
implement an iterative analysis workflow that begins with code understanding
and transitions to empirical performance analysis. This architecture offers
several advantages: it enables the use of a wider range of data sources than a
single prompt would allow, supports iterative refinement using outputs from
prior roles, and produces interpretable intermediate results. These
intermediate outputs can be reviewed to trace the origins of analysis claims.
In Section~\ref{sec:results}, we present the performance of \toolname against
baseline approaches as described in Sec.~\ref{sec:method:settings}.

Internally, the tool is organized into two iterative stages: the Source Code
Inspection Stage and the Profile Inspection Stage, each indicated by a yellow
box in the center of the diagram and expanded in detail above and below the
central diagram. These two stages are followed by an Analysis Aggregator step
and finally an Explanation Reviewer step. Below we explain and motivate each
stage, starting with the Source Code Inspection Stage and highlighting key
agent roles.

\subsection{Source Code Inspection Stage}

The Source Code Inspection Stage is responsible for extracting insight from the
kernel source code. It first prepares descriptions of the source files to
support selection, and then iteratively selects from the available code files
and refines a summary of the algorithm and a set of performance hypotheses
based on each code file. This stage repeats until all files have been reviewed
at least once, after which \toolname execution proceeds to the Profile
Inspection Stage.

A key component of this stage is the Performance Hypothesizer role, which
reviews the current selected code file and the current algorithm summary to
update a running set of performance hypotheses. These performance hypotheses
are the tool's heuristic predictions for how the kernel should perform based on
the code alone. They serve as a plan for later exploration of empirical
performance data.

\subsection{Profile Inspection Stage}

The Profile Inspection Stage is responsible for extracting insight from the NCU
profiles. It iteratively selects one or more profiles and a subset of metrics,
and produces a performance analysis for each selected group. It also optionally
leverages the DrGPU tool to update the analysis based on its suggestions. This
stage repeats until all profiles have been analyzed at least once, after which
\toolname execution continues to the Analysis Aggregator and Explanation
Reviewer steps. Several roles are notable in this stage, including:

\begin{itemize}
    \item \textbf{Profile Selector:} selects the next NCU profile(s) to analyze
    based on the algorithm summary, performance hypotheses, and any previous
    performance analyses. It attempts to choose profiles that will maximize the
    information gained from the next analysis pass, particularly as related to
    the performance hypotheses.
    \item \textbf{Metric Selector:} selects a subset of NCU metrics to analyze
    next from those available across all chosen profiles. This role reduces the
    noise in the analysis by focusing on the most relevant metrics for the
    current pass.
    \item \textbf{Profile Analyzer:} examines the profile metrics, algorithm
    summary, source code, and analysis guidelines
    (Section~\ref{sec:tool:performance-analysis-guidelines}) to produce a
    detailed performance analysis. This role's prompt requests analysis of
    performance bottlenecks, how tuning knobs impact performance, and
    suggestions to improve performance. It also requests inline citations to
    the Nsight Compute profiles referenced in the analysis to mitigate
    hallucination risk and support human review.
    \item \textbf{DrGPU Evaluator:} invokes DrGPU on the profile data and
    incorporates any suggestions from DrGPU it determines are useful. This role
    is optional, and we evaluate \toolname with and without it in
    Section~\ref{sec:results}.
\end{itemize}

\subsection{Aggregation and Review}

The Analysis Aggregator and Explanation Reviewer roles are responsible for
preparing the two outputs of \toolname. The Aggregator role combines all the
performance analyses generated for each Profile Inspection Stage pass into a
final performance explanation report. The Reviewer role compares the final
performance explanation against the performance hypotheses generated before
viewing any profile data to provide a final assessment of each hypothesis --
whether it is confirmed, refuted, or inconclusive given the generated
performance analysis. The reviewer role is optional and provides additional
interpretability for the explanation generation process.

\subsection{Performance Analysis Guidelines}
\label{sec:tool:performance-analysis-guidelines}

The Performance Analysis Guidelines are a set of general human-written GPU
performance analysis guidelines included as context for the Profile Analyzer
and Aggregator agents. We constructed these guidelines as a concise summary of
expert wisdom about GPU performance analysis. They are distilled from NVIDIA
documentation, existing literature on best practices, and consultation with
expert practitioners.

In contrast to prior tools~\cite{hao2023drgpu,zhou2021gpa}, \toolname leverages
LLMs to combine hypothesis-driven analysis, adaptive metric and profile
selection, and multi-pass analysis aggregation to support scalable,
interpretable, and algorithm- and architecture-aware interpretation of GPU
performance data. As with all LLM-based tools, \toolname may produce
lower-quality output depending on the quality of the LLM used and the quantity
of profiling data provided. We evaluate these trade-offs empirically in our
ablation study described in Section~\ref{sec:ablation-studies}.

%% file: method.tex
In this section, we describe our methodology for evaluating \toolname,
including other methods we compare with, chosen test cases, downstream tasks
and metrics, and the ablation study design. We also describe the LLMs and
hardware used.

\subsection{\toolname and Other Methods used for Comparison}
\label{sec:method:settings}

We compare two variations of \toolname against four other methods for
understanding GPU kernel performance data:
\begin{itemize}
    \item \textbf{Code Only}: The method does not generate a performance
explanation using profile data. We simply provide the kernel source code to the
LLM for the downstream task.
    \item \textbf{Code+Data}: A simple report containing only the raw metric
data table extracted from NCU metric data and the kernel source code is
provided to the LLM for the downstream task.
    \item \textbf{DrGPU Only}: We use DrGPU to generate a report of the kernel
performance, statically converting the suggestions and tree structure into a
basic natural language report without using LLMs.
    \item \textbf{LLM+DrGPU}: We pass the textified DrGPU report to a basic LLM
prompt including the raw NCU data and kernel source code to generate a
performance report.
    \item \textbf{\toolname Only}: We use \toolname to generate a performance
explanation report of the kernel, with the DrGPU Suggestion Reviewer agent
disabled.
    \item \textbf{\toolname{}+DrGPU}: We use \toolname with the DrGPU
Suggestion Reviewer agent enabled, meaning that \toolname evaluates the
suggestions generated by DrGPU and incorporates them into the report if deemed
helpful.
\end{itemize}

Each of these methods generates output, whether
basic code, or data tables, or a natural language report, which we then provide
as context to an LLM to complete one of two downstream tasks described below in
Section~\ref{sec:method:downstream-tasks}.

\begin{table*}[ht]
    \centering
    \caption{Evaluation cases used to evaluate \toolname and baseline approaches.}
    \label{tab:evaluation-cases}
    \begin{tabular}{lllllcc}
        \toprule
        \textbf{Application} & \textbf{Kernel}                                                 & \textbf{Algorithmic Motif} & \textbf{Scientific Domain} & \textbf{SLoC} & \textbf{DrGPU?} & \textbf{GPA?} \\
        \midrule
        b+tree (r)           & findRangeK                                                      & Graph Traversal            & Search                     & 62            & \textbullet     & \textbullet \\
        backprop (r)         & bpnn\_layerforward\_CUDA                                        & Unstructured Grid          & Machine Learning           & 67            & \textbullet     & \textbullet \\
        pathfinder (r)       & dynproc\_kernel                                                 & Dynamic Programming        & Grid Traversal             & 81            & \textbullet     & \textbullet \\
        nw (r)               & needle\_cuda\_shared\_1                                         & Dynamic Programming        & Bioinformatics             & 99            & \textbullet     & \textbullet \\
        hotspot (r)          & calculate\_temp                                                 & Structured Grid            & Material Science           & 113           & \textbullet     & \textbullet \\
        huffman (r)          & vlc\_encode\_kernel\_sm64huff                                   & Finite State Machine       & Lossless Compression       & 124           & \textbullet     & \textbullet \\
        lavaMD (r)           & kernel\_gpu\_cuda                                               & N-body Simulation          & Molecular Dynamics         & 210           & \textbullet     & \textbullet \\
        heartwall (r)        & kernel                                                          & Structured Grid            & Medical Imaging            & 1327          & \textbullet     & \textbullet \\
        LULESH               & \makecell[l]{ApplyMaterialPropertiesAnd-\\UpdateVolume\_kernel} & Unstructured Grid          & Hydrodynamics              & 392           & \textbullet     & \\
        XSBench              & xs\_lookup\_kernel\_baseline                                    & Monte Carlo                & Nuclear Physics            & 380           &                 & \\
        gaussian (r)         & Fan2                                                            & Dense Linear Algebra       & Linear Algebra             & 17            &                 & \textbullet \\
        \bottomrule
    \end{tabular}
\end{table*}

\subsection{Benchmarks and Applications used for Evaluation}
\label{sec:evaluation-cases}

We select a suite of benchmarks and proxy applications to evaluate the quality
of \toolname's analysis of GPU kernels, including some reused from prior
work~\cite{zhou2021gpa, hao2023drgpu}, as well as others of our own selection.
Below, we briefly describe the benchmark suite and proxy applications used in 
this study:
\begin{itemize}
    \item \textbf{Rodinia} is a benchmark suite of GPU kernels covering a wide
    range of algorithms and applications. We select nine kernels from Rodinia.
    \item \textbf{LULESH} is a hydrodynamics proxy application, which solves a
    Sedov blast wave problem using unstructured mesh data structures. We use
    version 2.0 of LULESH.
    \item \textbf{XSBench} is a proxy application for OpenMC, a Monte Carlo
    neutron transport code. Its single kernel represents the macroscopic
    cross-section lookup calculation, which we test with a variety of kernel
    launch parameter settings.
\end{itemize}

Table~\ref{tab:evaluation-cases} summarizes the applications and their
respective kernels selected for this study. For each application, we list the
name of the kernel profiled, whether the kernel was studied in the DrGPU or GPA
efforts~\cite{zhou2021gpa, hao2023drgpu}, as well as the algorithmic motif and
scientific domain represented by the application and kernel. As listed in
Table~\ref{tab:evaluation-cases}, these kernels cover a range of algorithmic
motifs and scientific domains while maintaining overlap with prior work for
ease of comparison. For each application, we select the kernel that the
application spends the most time in. When a kernel is invoked multiple times by
the application, we select the longest-running instance to profile. When
multiple instances take up the most time, we select the instance with the
median invocation time.

\subsection{Downstream Tasks and Metrics}
\label{sec:method:downstream-tasks}

We quantitatively evaluate the quality of \toolname outputs against the other
methods described above using two downstream tasks. For each task, we provide
the tool or comparison method's report as context to an LLM and ask it to solve
a related problem using the provided report as context. If an LLM achieves
higher downstream task performance when using one method's report as context
compared to another, we conclude that the former method's report is more
helpful.

\vspace{0.08in}
\noindent\textbf{Task 1: Multiple-Choice Question Answering (MCQ)}:
The multiple-choice question answering (MCQ) task asks the downstream LLM to
answer a set of multiple-choice questions about the performance of the kernel,
using the generated report (if provided) as context. These questions, twenty
per kernel, are manually written to assess understanding of the kernel code and
NCU profile data. For MCQ we estimate score@1, the expected percentage score on
the MCQ test given one attempt, based on scores achieved with twenty attempts
for each report generated. The score@1 is equivalent to the average percentage
score across all twenty attempts. We provide an example of an MCQ question in
Listing~\ref{lst:mcq-example}. We evaluate MCQ on a subset of applications due
to the manual effort required to write the questions.

\begin{lstlisting}[caption={Example MCQ question}, label={lst:mcq-example}]
"question": "Which of these hardware limits is most heavily saturated by this kernel?",
"correct_choices": ["SM issue rate"],
"incorrect_choices": ["ADU pipeline throughput",
                      "ALU pipeline throughput",
                      "CBU pipeline throughput"]
\end{lstlisting}

\vspace{0.08in}
\noindent\textbf{Task 2: Code Optimization (OPT)}:
The code optimization (OPT) task asks the downstream LLM to implement the
suggestions for code optimization provided in the report to maximize the
performance of the kernel. If the report does not provide suggestions for code
optimization, we prompt the LLM to identify optimizations itself as part of the
OPT task. When updated kernel code does not compile or pass test cases, we
provide this feedback to the LLM and ask it to try again, up to a fixed limit
on retries. More details are provided in
Section~\ref{sec:setup:downstream-tasks}. To consider the rate of valid and
invalid code generation, we also report pass@1, the expected percentage of
attempts that generate correct code.

\subsection{Ablation Studies}

To motivate the design of \toolname and choices made in the final downstream
task evaluation, we perform several ablation studies. Specifically, we study
the impact of the two key agent roles (metric selector and profile selector),
as well as the impact of number of profiles provided to the tool and the choice
of LLM used in \toolname. These studies are described in detail in
Section~\ref{sec:ablation-studies}.

%% file: setup.tex
In this section, we describe the setup for our experiments evaluating \toolname
outputs.

\subsection{Hardware Used}

We use the NVIDIA H100 SXM5 as the primary hardware for this study. For the
LULESH multi-profile ablation study (Sec.~\ref{sec:ablation-studies}), we also
include NVIDIA V100 and A100 GPUs, to represent a range of GPU architecture
generations.

\subsection{Application Configuration and Profile Collection}

For all applications, we use default input parameters to check correctness,
generate profiles to analyze, and profile after optimizations. All profiles are
collected with all NCU metric sections enabled along with metrics needed for
DrGPU analysis. We also manually export the line-level data to support DrGPU
analysis in both baseline and \toolname evaluations.

For the main evaluation, we collect and provide to the tool only one
representative profile for each kernel, collected using the default settings
and input. For ablation studies (Section~\ref{sec:ablation-studies}), we
collect and provide to the tool multiple profiles for each kernel, collected
using a range of performance knob settings and in some cases multiple GPU
architectures.

\subsection{Report Generation and OPT Task Setup}
\label{sec:setup:downstream-tasks}

For LLM-based report generation methods, we generate three independent reports
per experimental setting to account for stochasticity in LLM output. For
deterministic methods (Code Only, Code+Data, DrGPU Only), we generate a single
report. Each generated report is evaluated using the downstream task LLM,
gpt-oss-120b, over twenty independent attempts to estimate expected performance
with a single attempt using speedup@1, pass@1, and score@1 metrics. We use the
same LLM (gpt-oss-120b) to complete all downstream tasks to ensure fair
comparisons.

For the OPT task, we allow up to three retries maximum per OPT solving attempt
to balance solution recovery and inference costs. We measure kernel performance
using Nsight Systems (NSYS) to extract exact kernel execution time, recording
the average execution time over three runs per OPT solving attempt to minimize
the impact of any performance variability. We use NSYS rather than Nsight
Compute for this measurement to minimize profiler overhead in speedup
measurement, and we configure it to collect only minimal kernel timing
data. For OPT we estimate speedup@1, the expected speedup in performance over
the unoptimized kernel code achieved with one attempt, based on speedups
achieved with twenty attempts for each report generated. Attempts that fail to
generate valid code after three retries are excluded from the speedup@1
estimate.

\subsection{Setup for Ablation Studies}
\label{sec:ablation-studies}

We perform four ablation studies to understand the impact of configuration
choices and number of profiles provided on \toolname's OPT performance. These
are described in detail below.

In multi-profile configurations, we study ablation settings using only the
XSBench and LULESH applications, both to constrain inference costs and because
these applications have readily available tuning knobs. For LULESH we use up to
48 profiles, varying GPU architecture, block size, and maximum number of
registers per thread; for XSBench we use up to 75 profiles, varying grid type
(unionized, hash, or nuclide), block size, and maximum number of registers per
thread. Single-profile settings use the same profiles and applications as the
main downstream task evaluations.

\subsubsection{Ablation Study 1: Metric Selector Agent}

We first study the impact of the use of the Metric Selector agent on \toolname
output quality, comparing OPT performance in two settings: (1) Metric Selector
agent enabled, (2) Metric Selector agent disabled. All other agent roles are
enabled for this study. We present results for both single-profile \toolname
runs and additionally discuss the results for multi-profile \toolname runs.

\subsubsection{Ablation Study 2: Profile Selector Agent}

We next study the impact of the use of the Profile Selector agent on \toolname
output quality, comparing OPT performance in two settings: (1) Profile Selector
agent enabled, (2) Profile Selector agent disabled. All other agent roles are
enabled for this study, and we present results only for multi-profile \toolname
runs, as the profile selector is not used in single-profile runs.

\subsubsection{Ablation Study 3: Profile Count}
\label{sec:method:ablation-study-3}

We further study the impact of the number of profiles provided to the tool on
OPT performance. We generate \toolname outputs with a range of profile counts
from one profile to all profiles available. All other agent roles are enabled
for this study. We select the profiles to use in each setting by prioritizing
profiles with knob settings closer to the defaults, and for LULESH profiles
from older GPUs. We rank profiles by a distance score from default
configurations. For each profile with a different configuration setting, we
compute a normalized distance between its setting and the default setting,
breaking ties lexicographically by filename. For XSBench, the default
configuration is the unionized grid type, 128 threads per block, and 64
registers per thread. For LULESH, the default configuration is the V100 GPU
architecture, 128 threads per block, and 64 registers per thread.

\subsubsection{Ablation Study 4: LLM Choice}

Finally, we study the impact of the choice of LLM used in \toolname on OPT
performance in the single-profile setting. Table~\ref{tab:llms} summarizes the
LLMs used. We test our tool with two open source LLMs as well as one paid API
LLM. These LLMs represent the state of the art in paid API models and open
source models while having reasonable inference costs. We include the larger
120B parameter gpt-oss and the smaller 30B parameter Nemotron-3 to evaluate the
impact of model size. We set reasoning effort to the highest setting for all
experiments.

\begin{table}[ht]
    \centering
    \caption{LLMs used to evaluate \toolname and baselines. Note that parameter
    counts for the closed-source GPT 5.1 are not published by OpenAI.}
    \label{tab:llms}
    \begin{tabular}{llr}
        \toprule
        \textbf{LLM}      & \textbf{Open Source?} & \textbf{Model Size} \\
        \midrule
        GPT-5.1           & No  & Unpublished        \\
        gpt-oss-120b      & Yes & 120B (5.1B active) \\
        Nemotron-3-Nano   & Yes & 30B (3B active)    \\
        \bottomrule
    \end{tabular}
\end{table}

%% file: results.tex
In this section, we present the results of the evaluation of \toolname. We
begin by presenting a brief snippet of \toolname's output for a specific
kernel. After describing the layout of results figures, we present results from
the four ablation studies, followed by evaluation of the tool's performance on
the downstream tasks of LLM multiple-choice question answering (MCQ) and code
optimization (OPT) for single-profile cases. Finally, we present the best
baseline and \toolname optimization techniques applied for each application.

\subsection{Sample \toolname Output}

We present a brief snippet of \toolname's output for the gaussian Fan2 kernel
in Listing~\ref{lst:example-output}. Some metric names are abridged for
brevity. After presenting this summary section, this particular report
continues with suggestions to address the latency and coalescing bottlenecks by
increasing the block size, refactoring the memory layout, and reducing grid
size where appropriate, enabling the downstream LLM to generate a kernel with a
14.0x speedup over the original code.

\begin{lstlisting}[caption={Example \toolname output for the gaussian Fan2 kernel}, label={lst:example-output}]
## 7. Summary of main bottlenecks

1. **Memory-latency bound, not bandwidth bound**
    - Long Scoreboard stalls dominate (`21.39` per issue).
    - L2 hit rate is high; `dram__throughput` is only 1.77%

2. **Very poor global memory coalescing**
    - `smsp__..._bytes_per_sector_mem_global_op_ld = 8.34 B/sector` (ideal is 32 B/sector).
    - `derived__..._sectors_global_excessive = 374,514` sectors.

3. **Low warp and thread-level utilization**
    - Blocks have 16 threads (half-warp).
    - Average active threads per issued inst ~= 15.15 (~50%
    - `sm__warps_active` only 10.4%

4. **Algorithmic tail and wasted threads**
    - Fixed grid size independent of pivot index `t` leads to many threads/blocks doing no useful work for larger `t`.
\end{lstlisting}

\subsection{Layout of Figures}

In all figures we refer to approaches by shortened names listed in
Section~\ref{sec:method:settings}. For speedup figures, we plot both the mean
speedup@1 score and the maximum speedup observed across all twenty attempts.
These are indicated by an X and a dash, respectively. We focus primarily on the
maximum speedup in our discussion, as it more closely reflects the speedup a
user would use in practice after generating twenty attempts. We include the
average speedup@1 to represent the overall performance tendency across all
attempts. All figures presenting per-application quantities are sorted by the
number of source lines of code in the kernel (with gaussian placed last to
allow for a separate y-axis). For brevity, we only present the best of the Code
Only and Code+Data methods as "Code-based" in all figures. For some ablation
study figures, we also present the harmonic mean~\cite{eeckhout2024rip} of the
speedup@1 scores across applications to understand the overall impact of
ablated roles.

\subsection{Ablation Study 1: Metric Selector Agent}

\begin{figure}[!ht]
    \centering
    \includegraphics[width=.49\textwidth]{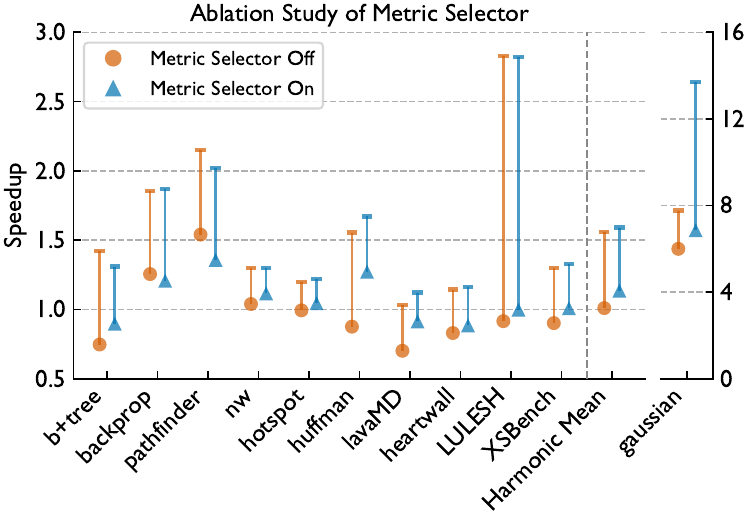}
    \caption{Average speedup@1 score and maximum speedup observed in
    single-profile cases with the Metric Selector agent enabled and disabled.}
    \label{fig:metric-ablation-single-profile-speedup}
\end{figure}

We present mean speedup@1 and maximum speedup observed in single-profile cases
with the Metric Selector agent enabled and disabled in
Figure~\ref{fig:metric-ablation-single-profile-speedup}. We observe that the
Metric Selector agent increases the maximum speedup observed for all
applications except b+tree, pathfinder, and LULESH, and similarly affects the
mean speedup@1. Overall, as reflected in the harmonic mean across applications,
the Metric Selector agent improves both mean and maximum speedups.

We also separately evaluate the impact of the Metric Selector agent in
multi-profile cases, finding that the Metric Selector agent has minimal impact
in those cases. Overall, given these results, we conclude that the Metric
Selector agent is more often than not beneficial, and we default to enabling
it.

\begin{figure}[!ht]
    \centering
    \begin{subfigure}[b]{0.205\textwidth}
        \centering
        \includegraphics[width=\textwidth]{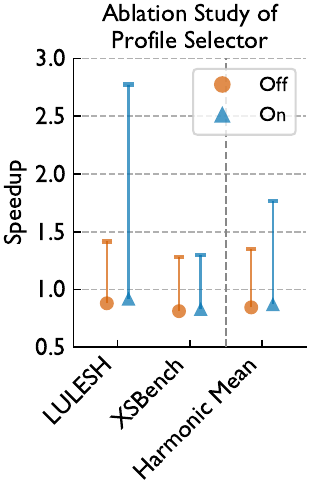}
    \end{subfigure}
    \hfill
    \begin{subfigure}[b]{0.275\textwidth}
        \centering
        \includegraphics[width=\textwidth]{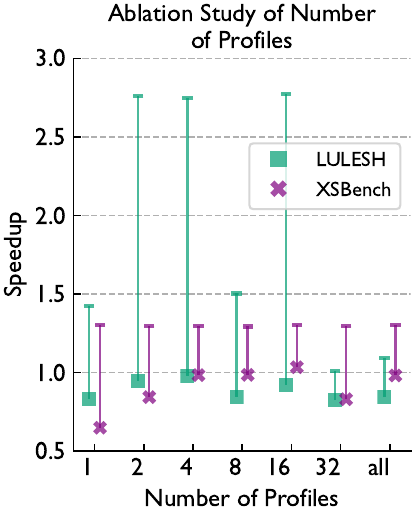}
    \end{subfigure}
    \caption{Left: Average speedup@1 score and maximum speedup observed in
    multi-profile cases with the Profile Selector agent enabled and disabled.
    Right: Average speedup@1 score and maximum speedup observed in
    multi-profile cases with varying profile counts from one profile to all
    profiles.}
    \label{fig:profile-ablation-multi-profile-speedup}
\end{figure}

\subsection{Ablation Study 2: Profile Selector Agent}

We present mean speedup@1 and maximum speedup in multi-profile cases with the
Profile Selector agent enabled and disabled in
Figure~\ref{fig:profile-ablation-multi-profile-speedup} (left). We observe that
the Profile Selector agent increases both mean and maximum speedup,
particularly maximum speedup observed for LULESH. As such, we default to
enabling the Profile Selector agent in \toolname.

\subsection{Ablation Study 3: Profile Count}

We present average speedup@1 score and maximum speedup observed in
multi-profile cases with varying profile counts from one profile to all
profiles in Figure~\ref{fig:profile-ablation-multi-profile-speedup} (right). We
observe a sweet spot effect in the number of profiles, with the optimal number
around four profiles for LULESH and sixteen profiles for XSBench. \toolname
benefits from additional profiles, but at significantly higher profile counts,
we infer that the additional profiles do not provide useful insight and degrade
performance by adding noise to the analysis context.

\subsection{Ablation Study 4: LLM Choice}

\begin{figure}[!ht]
    \centering
    \includegraphics[width=0.48\textwidth]{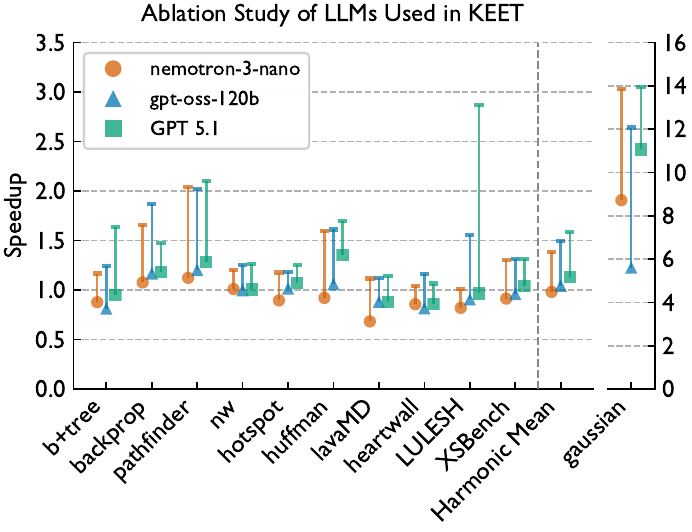}
    \caption{Speedup@1 by application and LLM for \toolname.}
    \label{fig:opt-keet-speedup}
\end{figure}

We break down speedup@1 averages and maximum speedups by LLM in
Figure~\ref{fig:opt-keet-speedup}, focusing on \toolname (without DrGPU)
experiments. Overall, GPT 5.1, the largest model, most often achieves the
highest average speedup@1 scores, followed by gpt-oss-120b. We note that
\toolname is able to generate useful reports across LLMs tested, using large
and small open-source LLMs as well as closed-source LLMs. However, the most
effective LLM is hard to predict for a particular application. For
\toolname{}+DrGPU experiments (plots not shown), we find that nemotron-3-nano,
the smallest model, most often achieves the highest mean speedup@1 scores,
suggesting that DrGPU suggestions are particularly effective in increasing the
performance of smaller LLMs. Given these results, we present GPT 5.1 results
for \toolname experiments and nemotron-3-nano results for \toolname{}+DrGPU
experiments in our final downstream task evaluation.

\subsection{Downstream Task 1: Multiple-Choice Questions (MCQ)}

\begin{figure}[!ht]
    \centering
    \includegraphics[width=0.48\textwidth]{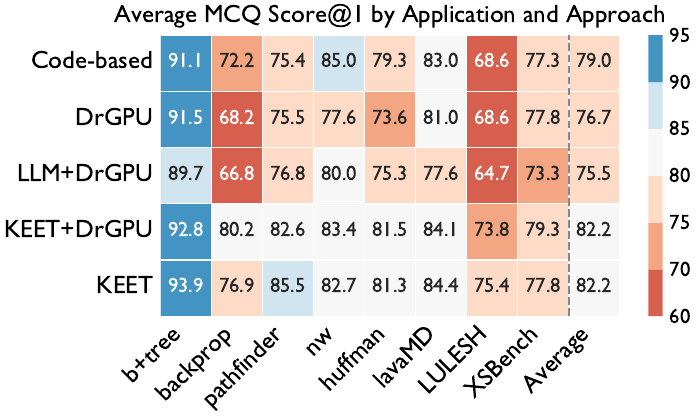}
    \caption{Average score@1 (out of 100) for the MCQ task grouped by
    application and context types.}
    \label{fig:mcq-overall-pass}
\end{figure}

The first downstream task we evaluate is multiple-choice question answering. We
present average score@1 values for the MCQ task grouped by application and
context types in Figure~\ref{fig:mcq-overall-pass}. Score@1 indicates the
estimated score out of 100 that the LLM should achieve on the MCQ task with one
attempt.

We observe that both \toolname and \toolname{}+DrGPU consistently achieve
higher score@1 scores than baseline methods across applications tested. The
exceptions are nw, where the Code-based method is able to achieve a higher
score@1 than both \toolname{}-based methods, and XSBench, where DrGPU ties with
\toolname{}. The backprop and LULESH application MCQ sets are particularly
difficult across all context types, and are areas where the \toolname methods
provide the greatest improvement over the baseline methods. Across
applications, \toolname and \toolname{}+DrGPU both achieve an average score@1
of 82\%, while the nearest baseline, Code+Data, achieves 79\%.

\begin{figure}[!ht]
    \centering
    \includegraphics[width=0.48\textwidth]{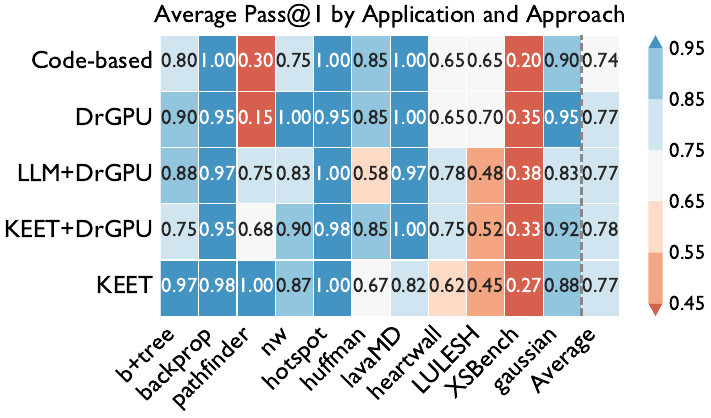}
    \caption{Pass@1 by application and approach.}
    \label{fig:opt-overall-pass}
\end{figure}

\begin{figure*}[!ht]
    \centering
    \includegraphics[width=0.7\textwidth]{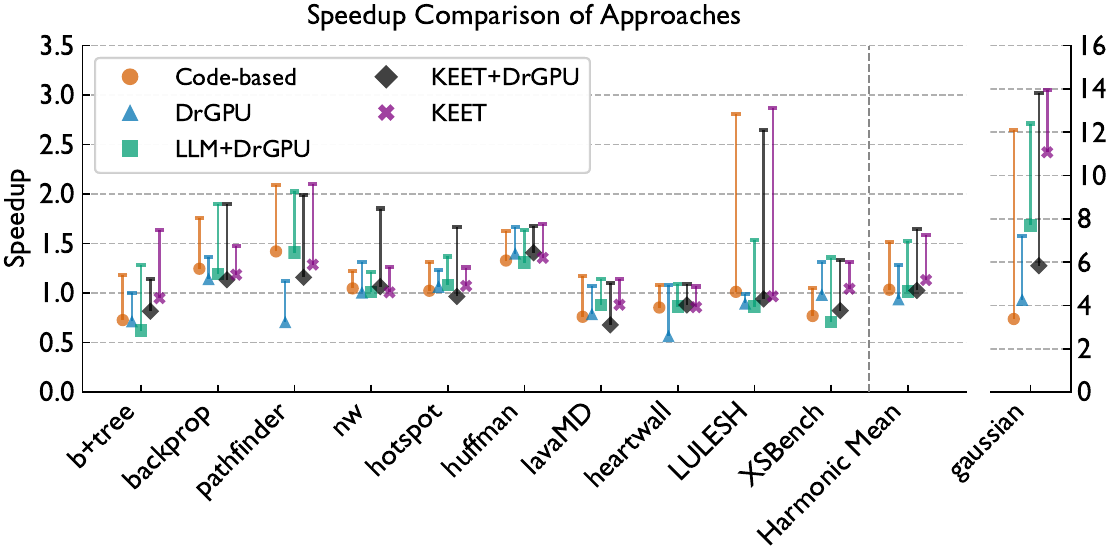}
    \caption{Speedup@1 by application and approach.}
    \label{fig:opt-overall-speedup}
\end{figure*}

\subsection{Downstream Task 2: Code Optimization (OPT)}

For OPT, we first present pass@1 scores grouped by application and context
configuration in Figure~\ref{fig:opt-overall-pass} to assess the difficulty of
maintaining correctness in the optimization task. Pass@1 indicates the
estimated likelihood of the LLM generating valid (compilable and test
case-passing) code in its optimization attempt. These scores vary significantly
across applications. The pathfinder, heartwall, LULESH, and XSBench
applications are particularly difficult for most methods to generate code for.

\begin{table*}[!ht]
    \centering
    \renewcommand{\arraystretch}{1.2}
    \begin{tabular}{lp{.37\textwidth}r|p{.37\textwidth}r}
        \toprule
        Application & Best baseline technique                                                                                              & S/up & Best \toolname technique                                                                            & S/up \\
        \midrule
        b+tree      & Use warp primitives, reduce block size, \textcolor{red}{stride loop}                                                 & 1.28 & Use warp primitives, reduce block size                                                              & \textbf{1.64} \\
        backprop    & Remove barriers, strength reduction, use shared memory                                                               & 1.90 & Remove barriers, remove shared memory                                                               & 1.90 \\
        pathfinder  & Add \_\_restrict\_\_, double-buffer shared memory, remove redundant calculation, use \_\_ldg                         & 2.09 & Add \_\_restrict\_\_, double-buffer shared memory, remove redundant calculation                     & 2.10 \\
        nw          & Inline helper, pad shared memory, \textcolor{red}{add \_\_restrict\_\_, use syncwarp, unroll loop, set cache config} & 1.32 & Inline helper, \textbf{reduce block size}, pad shared memory, \textbf{reduce barriers, use \_\_ldg} & \textbf{1.85} \\
        hotspot     & Reduce shared memory usage, \textcolor{red}{pad shared memory, add \_\_restrict\_\_}                                 & 1.37 & Reduce shared memory usage, \textbf{remove barriers, hoist invariants}                              & \textbf{1.67} \\
        huffman     & Use warp primitives, use shared memory, remove barriers                                                              & 1.66 & Use CUB scan, remove barriers, add macro helper                                                     & 1.70 \\
        lavaMD      & Add \_\_restrict\_\_, increase parallelism, use register accumulators, unroll loop                                   & 1.18 & Hoist invariant, use register accumulators, remove barriers, unroll loop                            & 1.14 \\
        heartwall   & Add \_\_restrict\_\_, unroll loop                                                                                    & 1.09 & Add \_\_restrict\_\_, add const, unroll loop, set cache config                                      & 1.09 \\
        LULESH      & Remove redundant calculation, \textcolor{red}{hoist loads}                                                           & 2.81 & Remove redundant calculation, \textbf{use \_\_ldg}                                                  & \textbf{2.87} \\
        XSBench     & Replace division with fast reciprocal, use \_\_ldg                                                                   & 1.36 & Hoist loads to registers, force inline helpers, unroll loops                                        & 1.33 \\
        gaussian    & Increase block size, change memory layout, \textcolor{red}{use shared memory}                                        & 12.4 & Increase block size, change memory layout, \textbf{reduce grid size, hoist global load}             & \textbf{14.0} \\
        \bottomrule
    \end{tabular}
    \caption{Best baseline and \toolname optimization techniques applied for
    each application. Techniques with speedup marked in bold are best by a
    significant margin (five percentage points or more) over the other
    technique listed for that application. For cases where one technique is
    significantly better than the other, differing components between the two
    techniques are highlighted with red (removed components) and bold (added
    components) text.}
    \label{tab:best-baseline-and-keet-techniques}
\end{table*}

Figure~\ref{fig:opt-overall-speedup} presents harmonic mean speedup@1 scores
and maximum speedup by application and approach. Speedup@1 indicates the
expected speedup in performance over the original (unoptimized) code achieved
with one attempt, with the harmonic mean taken over the speedup@1 scores across
all three runs for the LLM based approaches (LLM+DrGPU, KEET, KEET+DrGPU, see
Section~\ref{sec:setup:downstream-tasks}). As with pass@1, these scores vary
significantly across applications.

Across applications, \toolname achieves the highest maximum speedup in five
cases, and \toolname{}+DrGPU achieves the highest maximum speedup in three
cases. For lavaMD, the code-based method achieves the highest maximum speedup,
while for XSBench, the LLM+DrGPU method achieves the highest maximum speedup.
For heartwall, there is no significant difference in maximum speedup across
approaches. In all cases where a \toolname{}-based method does not achieve the
highest maximum speedup, the margin between the highest baseline and highest
\toolname{}-based method is less than five percentage points. Given these
results, we conclude that the most effective overall approach is \toolname,
with \toolname{}+DrGPU offering additional benefits for nw and hotspot.

\subsection{Reviewing Optimizations Applied}

To understand more specifically what techniques the best optimization attempts
apply to achieve strong speedups, we present the best baseline and \toolname
optimization techniques applied for each application in
Table~\ref{tab:best-baseline-and-keet-techniques}. These were identified based
on the optimizing LLM's own report on what techniques were applied as well as
human review of the code diffs. Techniques that achieve at least a five
percentage point greater speedup than the next best technique are marked in
bold. Frequently, the best baseline and \toolname technique sets have some
overlap, but the \toolname techniques that are most successful are more likely
to use \verb|__ldg| instructions and less likely to introduce new shared memory
usage. In some cases, the speedup improvement from \toolname techniques may be
due to a performance regression introduced by the baseline technique, as is the
case for b+tree. For nw and hotspot, where \toolname{}+DrGPU is most effective,
we observe that both technique sets include removing barriers as an improvement
over the baseline, suggesting that DrGPU is particularly helpful where
\toolname needs encouragement to identify unnecessary barriers. Overall, we
argue that \toolname is able to more effectively tune its optimization
suggestions to the specific performance profile data observed for the kernel
being optimized.

\begin{figure}[!ht]
    \centering
    \includegraphics[width=0.44\textwidth]{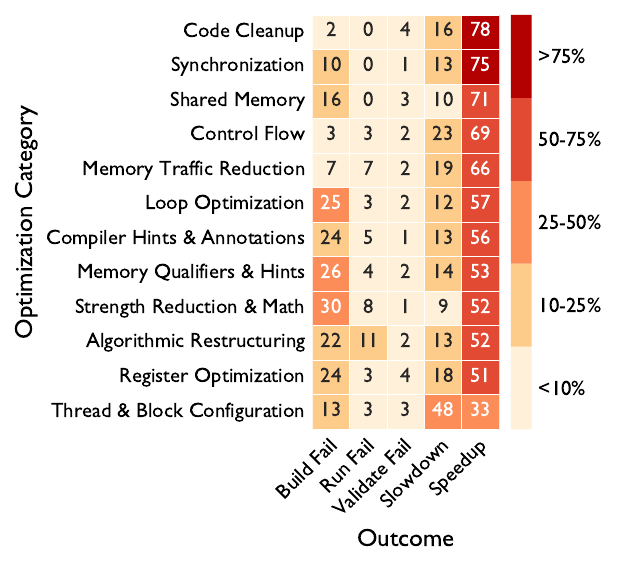}
    \caption{Heatmap of percentage of attempts in each outcome category that use each optimization technique.}
    \label{fig:taxonomy-technique-outcome}
\end{figure}

We also systematically analyze the optimization techniques applied against
attempt outcomes across all applications and approaches, presenting the results
in Figure~\ref{fig:taxonomy-technique-outcome}. These optimization technique
labels are automatically generated by post-processing the optimization attempt
code diffs and LLM post-optimization reports using another LLM. We observe that
loop optimizations, memory qualifiers and hints, algorithmic restructuring, and
strength reduction and math changes are most associated with build and run
failures. Notably, thread and block configuration changes have significantly
greater slowdown rates than any other optimization technique -- most likely
because it is generally difficult to tune block and grid size parameters based
on a single profile.

%% file: related.tex
Existing non-LLM-based tools for analyzing GPU kernel performance data include
GPA~\cite{zhou2021gpa, zhou2021automated} and DrGPU~\cite{hao2023drgpu}. GPA,
the GPU Performance Advisor, is a standalone tool that combines static analysis
and runtime stall measurement to attribute stalls to instructions and suggest
optimizations based on performance models~\cite{zhou2021gpa}. DrGPU, which
succeeds GPA and presents results in comparison to GPA, is a tree-centric tool
which analyzes stall reasons collected with NCU and decomposes them into a tree
structure, offering suggestions to address the most frequent stall
reasons~\cite{hao2023drgpu}. We discuss DrGPU in greater detail in
Section~\ref{sec:background:drgpu}, as \toolname can optionally integrate
DrGPU's suggestions into its analysis. We also directly compare \toolname
against DrGPU in our evaluation. While both of these tools are useful for
interpreting GPU kernel performance data, they are limited to providing generic
suggestions based on fixed metric thresholds or performance models. They also
cannot reason generally about kernel source code and its relationship to
performance bottlenecks.

There is also extensive prior work on non-LLM-based tools for automating
performance analysis in HPC contexts generally. Hatchet~\cite{bhatele:sc2019}
enables automating common performance analysis tasks in Python across profiling
tools. Pipit~\cite{bhatele:2023pipit} proposes a similar framework focusing on
execution trace analysis. HPCToolkit~\cite{hpctoolkit,adhianto2024refining} has
extensive support for collecting and visualizing performance profiles for both
CPU and GPU workloads across vendors. AMD's ROCm Compute
Profiler~\cite{rocmcompute} is a tool for collecting performance data for AMD
GPUs, similar to Nsight Compute for NVIDIA GPUs~\cite{ncu}.
PARAVER~\cite{paraver} is a tool for visualizing parallel code execution
traces. EasyView~\cite{zhao2024easyview} incorporates profile data directly
into integrated development environment (IDE) software. Keiff et
al.~\cite{keiff2022automated} propose a framework that provides a unified
interface for a range of performance analysis tools and simplifies their usage.
HTA~\cite{hta} supports analysis of PyTorch application performance, and
Scalasca~\cite{geimer2010scalasca} automatically analyzes execution traces for
specific patterns and provides an interactive report and timeline to explore
the results. CATS~\cite{schaad2025cats} supports tracing and analysis of memory
and control flow patterns in parallel programs to identify performance
bottlenecks. \toolname focuses specifically on the problem of interpreting
low-level GPU performance metrics, and leverages LLMs to provide natural
language explanations and optimization suggestions tuned to the specific kernel
under analysis.

LLMs have been shown to be useful for generating, reasoning about, and
optimizing kernel source code, with and without performance data. Nichols et
al.~\cite{nichols:hpdc2024} develop ParEval, a framework which tests LLM
capabilities for generating parallel code, including CUDA kernels. Cui et
al.~\cite{cui2025large} assess LLM understanding of performance optimization
using a suite of CPU-based HPC codes from varying domains, finding that LLMs
struggle with maintaining correctness when optimizing code. Lange et
al.~\cite{lange2025towards} propose a framework which uses LLMs to translate
PyTorch code into CUDA kernel code, and then optimize the kernel code in an
agentic loop including a profiling stage. Zaeed et al.~\cite{zaeed2025opal}
propose Opal, a framework which incorporates NCU metrics and rule engine
outputs to prompt GPT-4o to optimize CUDA kernels. Yu et
al.~\cite{yu2026towards} survey the wide range of proposed techniques for using
LLMs to generate CUDA kernels. Dai et al.~\cite{dai2026cuda} propose CUDA
Agent, an agentic RL framework incorporating LLMs to generate CUDA kernels in
an agentic development environment with access to profiling tools. Zhang et
al.~\cite{zhang2025cudaforge} propose CudaForge, a framework incorporating NCU
data with Judge and Coder agents to optimize CUDA kernels. These frameworks
have shown the promise of using LLMs to generate and optimize CUDA kernels with
the assistance of profiling tools. Nsight Compute itself also includes NCU
Copilot, an LLM integrated into the GUI to provide explanations of GPU and
kernel performance concepts and suggest code changes~\cite{ncu}. \toolname
builds upon these efforts by focusing on the problem of generating natural
language explanations of GPU kernel performance data, specifically in an HPC
context.

%% file: conc.tex
We have presented \toolname, an LLM-based agentic framework for explaining GPU kernel
performance using data from Nsight Compute profiles. We evaluated the quality
of \toolname's analysis of GPU kernels against DrGPU and simple code-based
baseline approaches using two downstream tasks: multiple-choice question
answering and code optimization. This evaluation indicates that \toolname's
analysis of GPU kernels is effective in many cases at providing advice that
LLMs can use to provide better answers to multiple-choice questions and
implement superior optimizations to kernel code. We also find that \toolname
effectively integrates DrGPU's suggestions to deliver further speedup
improvements in some cases, suggesting future potential for integrating
additional new and existing performance analysis tools into the \toolname
framework. Finally, we presented the best baseline and \toolname optimization
techniques applied for each application, showing that \toolname can more
effectively tune optimization suggestions to empirical performance data,
avoiding performance regressions and improving final speedup achieved in
several cases.

Overall, we argue that \toolname is a valuable tool for generating kernel
performance explanation reports to support both LLM and human-in-the-loop
performance optimization and analysis of Nsight Compute profile data. Our
evaluation demonstrates that \toolname can enable low-effort, high-impact
LLM-based kernel performance optimization, including updating of existing CUDA
kernel code to maximize performance on newer GPU architectures. \toolname can
also gain insight from additional input profiles. Future effort in this
direction can benefit from our analysis of the specific optimization techniques
applied and their success rates. We specifically highlight the opportunity to
integrate existing work on performance knob autotuning to improve \toolname's
support for optimizing grid and block size parameters.